# Spin to charge conversion in MoS2 monolayer with spin pumping

Cheng Cheng[1,4*], Martin Collet[1], Juan-Carlos Rojas Sánchez[1], Viktoria Ivanovskaya[1], Bruno Dlubak[1], Pierre Seneor[1], Albert Fert[1], Hyun Kim[2,3], Gang Hee Han[2], Young Hee Lee[2,3], Heejun Yang[2,3], Abdelmadjid Anane[1*]

[1] Unité Mixte de Physique, CNRS, Thales, Univ Paris-Sud, Université Paris-Saclay, 91767, Palaiseau, France

[2] IBS Center for Integrated Nanostructure Physics (CINAP), Institute for Basic Science, Sungkyunkwan University, Suwon 440-746, Korea

[3] Department of Energy Science, Sungkyunkwan University, Suwon 440-746, Korea

[4] Department of Information Science and Electronic Engineering, Zhejiang University, Hangzhou 310027, China

*To whom correspondence should be addressed;
E-mail:     madjid.anane@u-psud.fr
            coracoracheng@outlook.com

**Layered transition-metal dichalcogenides (TMDs) family are gaining increasing importance due to their unique electronic band structures[1-4], promising interplay among light, valley (pseudospin), charge and spin degrees of freedom[5-10]. They possess large intrinsic spin-orbit interaction which make them most relevant for the emerging field of spin-orbitronics[11]. Here we report on the conversion of spin current to charge current in MoS$_2$ monolayer. Using spin pumping from a ferromagnetic layer (10 nm of cobalt) we find that the spin to charge conversion is highly efficient. Analysis in the frame of the inverse Rashba-Edelstein (RE) effect yields a RE length in excess of 4 nm at room temperature. Furthermore, owing to the semiconducting nature of MoS$_2$, it is found that back-gating allows electrical field control of the spin-relaxation rate of the MoS$_2$–metallic stack.**

Spintronics has started as basic questionings about the coupling between charge currents and spin currents in nanostructures which led the field to evolve rapidly and soon yielded practical devices such as tunnel magnetoresistance (TMR) read heads and magnetic random access memories (MRAMs). Many of the spintronics concepts including the yet elusive spin-transistor[12] rely on efficient and reciprocal, charge to spin signal conversion. This conversion has long been performed through magnetoresistive effects using metallic ferromagnets, however it is only recently that the spin-Hall effect (SHE) had made it possible to use non-magnetic materials to perform this conversion as long as they possess large spin orbit coupling (SOC). The figure of merit of SHE is given by the spin Hall angle ($\theta_{SH}$). For Pt, by far the most widely used SOC material, $\theta_{SH}$ ranges from 0.06 [13] to 0.19 [14] depending on the measurement method, but for β-W it can reach up to 0.33 [15]. SHE torques are most likely to be implemented in future spintronic devices like spin-orbit-torque-MRAMs[16]. As SHE torques are interfacial in nature, the bulk of the SOC metallic layer is of little use and could even be detrimental to the energy efficiency of the device. It is therefore appealing to consider true 2D SOC systems for the spin-current to charge-current conversion and vice versa. A few experiments have paved the way toward this direction. Graphene has been reported to possess small spin to charge conversion efficiency ($\theta_{SH} \sim 10^{-7}$ to $3.10^{-3}$) [17,18] unless SOC is extrinsically induced ($\theta_{SH} \sim 0.2$) [19,20]. Topological interfacial states have shown larger spin/charge conversion efficiencies, either at the Ag/Bi interface[21] or at the surface of Bi$_2$Se$_3$[22,23] and α-Sn[24]. However, in these cases the active 2D electronic states coexist with the bulk states, making it difficult to distinguish the contribution of the interface/surface from that of the more conventional 3D states[25,26]. In this report we address a prototypical 2D material: MoS$_2$ monolayer. MoS$_2$ belongs to TMDs family. Recently MoS$_2$ induced spin-transfer-torque has been shown[27] but the spin to charge conversion remains to be observed. To demonstrate the spin to charge conversion, we use the inverse RE effect configuration, originally proposed for SHE by Hirsch[28], where a spin current injected into a nonmagnetic (NM) material generates a transverse charge imbalance, detected as a voltage signal. Pure spin current is generated by spin pumping[29,30], where the magnetization of a thin Co film is set in precession under ferromagnetic resonance (FMR) conditions. The out-of-equilibrium spin accumulation in Co pumped by the rf-magnetic field diffuses to the adjacent NM layer (here MoS$_2$) which offers extra-spin-relaxation channels provided that the SOC is strong enough in this NM layer. Compared with spin-polarized charge current injection[31,32], the spin pumping approach[33] circumvents the impedance mismatch problem at the FM/semiconductor interface [34,35].

The experimental layout is illustrated in Figure 1. A n-Si/SiO$_2$ wafer (Fig. 1a), with its center area ( 5 mm × 5 mm) covered by chemical vapor deposited (CVD) monolayer MoS$_2$, is used as a substrate of a continuous ferromagnetic (FM) stack (Al 3 nm/Co 10 nm/Al 3 nm/Cu 3 nm) deposition. The cobalt layer has therefore two equivalent interfaces which rules out any Rashba-like contribution to the signal from within the FM[36]. Note here that the Al and Cu both have long spin diffusion length and do not act as spin sink. Samples are then cut so that the part with no MoS$_2$ flakes are used as control samples (*C*) while MoS$_2$\Al\Co\Al\Cu structures are hereafter labeled as *S* (Fig. 1b,c). As can be seen from the optical microscope image in Fig. 1d, the MoS$_2$ layer is not continuous but consists of isolated monolayer islands with lateral dimension of ~20 μm; the areal coverage on *S* is ~40% (see methods).

As in any standard spin-pumping experiment involving metallic ferromagnets, there are two contributions to the photo-voltage signal. One is the voltage of interest that is to be presumably, in our case, attributed to MoS$_2$. Here the MoS$_2$ islands act as sources of a dc charge current I$_C$, driven through the full stack. The other is the ubiquitous well-known spin-diode effect generated within the FM layer itself due to anisotropic magnetoresistance (AMR) and planar Hall effect (PHE)[37]. These two contributions can be distinguished according to their different lineshapes and angular dependences: while the AMR and the PHE voltages can have both dispersive (D) and Lorentzian (L) components, the spin to charge conversion gives a purely Lorentzian peak[38]. Al has specifically been chosen as the spacer between the Co layer and the MoS$_2$ monolayer to increase the transparency of the MoS$_2$\metal interface to spin current. There are two reasons for that : the first is that Al work function (~ 4 eV ) is the closest to the electron affinity of monolayer MoS$_2$ among light metals[39] (Fig. 1f). The "effective Schottky" barrier height is therefore minimized to ensure large diffusive spin flow through the Al/MoS$_2$ interface. Secondly, as stated earlier, Al is not a spin sink due to its low SOC.

Figure 2 presents the result of the voltage measurements on samples *C* (control) and *S* (FM stack on MoS$_2$). Fig. 2a and Fig. 2c are the measured photo-voltage at 8 GHz for *C* and *S*, respectively, under the FMR condition. For the control film *C*, the lineshape of the voltage signal is mostly dispersive (green curve, 1, D), highlighting the expected contribution from AMR and PHE. Note that the small Lorentzian component (blue curve, 1, L) does not change sign with reversed magnetic field H. For sample *S* which includes the discontinuous MoS$_2$ monolayer, two peaks are present in the measured photo-voltage signal shown as black circles in the upper panel of Fig. 2c. The fitted gray line includes a sum of two peaks: Peak 1 (green curve) appears at the same bias field as the peak in sample *C*. A new peak (red curve, 2) appears on the lower field side of Peak 1. While the green peak 1 does not have a significant Lorentzian component, the red peak 2 presents a strong Lorentzian signal, as shown in the upper panel of Fig. 2c by (the red cross marks). Because of the discontinuity of the MoS$_2$ monolayer (Fig. 1d) in sample *S*, we attribute the origin of peak 1 (green) to the FM layer directly in contact with the Si\SiO$_2$ substrate *i.e.* the gaps in between the MoS$_2$ islands, while we ascribe the additional peak (2, red) to the FM layer deposited on top of the MoS$_2$ islands. We use the fact that the resonance fields are different for the two FM regions to spectroscopically discriminate their photo-voltage contributions. We can therefore access simultaneously and under the same experimental conditions to the signal where no spin to charge conversion is expected (FM on Si/SiO$_2$) and the signal of interest (FM on MoS$_2$). The Lorentzian signal, as shown in the upper panel of Fig. 2c by (the red cross marks) changes sign when reversing the magnetic field, in contrast with the L component presented by the blue curve in Fig. 2a for *C*. The lineshape and the H field symmetry of peak 2 in sample *S* are the signature of the spin to charge conversion origin of this signal.

To further strengthen this analysis of the complex lineshapes measured on *S*, we performed a systematic study over the frequency range of 4-8 GHz with step of 0.2 GHz. The extracted linewidths of peaks 1 and 2 and their resonance field positions are presented in Fig. 2b for sample *C*, and in Fig. 2d,e for sample *S*. The fitting result of the green peak 1 in *S* (green squares in d and e) coincides well with the single peak in the reference sample *C* (green circles in b). The newly appeared red peak 2 in *S* always occurs on the lower field side of the initial green peak 1 and has a larger linewidth, as presented by the red circles in d and e. Cu\Al\Co\Al layers when deposited on SiO$_2$ and on MoS$_2$ have also distinctive effective magnetizations (see supplementary information) with the FM layer deposited on MoS$_2$ having the usual cobalt material parameters. This is due to different wetting properties of Al when grown on SiO$_2$ and on MoS$_2$. For instance, a single peak is observed when the Al spacer layer between Co and MoS$_2$ is missing.

After identifying the spin to charge conversion signal presented in Fig. 2c by the red cross marks we quantify the conversion rate following the calculations by Rojas Sánchez *et al.*[21]. The spin current density pumped from the Co 10 nm layer under FMR into the MoS$_2$ monolayer is

$$J_s = \frac{G_{\uparrow\downarrow}^{eff} \gamma^2 \hbar h_{rf}^2}{8\pi\alpha^2} \left[ \frac{4\pi M_{eff}\gamma + \sqrt{(4\pi M_{eff}\gamma)^2 + 4\omega^2}}{(4\pi M_{eff}\gamma)^2 + 4\omega^2} \right] \frac{2e}{\hbar} \qquad \text{(eq. 1)}$$

Where $h_{rf}$ is rf-magnetic field, γ is the gyromagnetic ratio, α is the Gilbert damping parameter and $M_{eff}$ is the Co effective magnetization obtained through FMR measurements (see supplementary information). For a given frequency $\omega/2\pi$, the main unknown quantity in (eq. 1) is the effective spin-mixing conductance $G_{\uparrow\downarrow}^{eff}$ at the MoS$_2$/(Al 3nm/Co 10nm) interface, given by

$$G_{\uparrow\downarrow}^{eff} = \frac{4\pi M_s t_F}{g_{eff} \mu_B} (\Delta\alpha) \qquad \text{(eq. 2)}$$

, where $M_s$ and $t_F$ are the Co magnetization and thickness, $g_{eff}$ is the effective Landé factor, $\mu_B$ is the Bohr magneton and $\Delta\alpha$ is the increase in the Gilbert damping constant due to spin pumping. To extract $G_{\uparrow\downarrow}^{eff}$, the usual procedure consists of comparing the Gilbert damping parameter between a reference ferromagnet (α$_0$) and the same ferromagnet capped with the spin-sink layer (α), attributing the total increase in damping $\Delta\alpha = \alpha - \alpha_0$ to the spin-pumping effect. Here this procedure breaks down since the cobalt grown on the spin-sink layer (MoS$_2$) has different material parameters than the one grown on SiO$_2$. We can however still calculate an upper bound for $G_{\uparrow\downarrow}^{eff}$ of 1.54 × 10$^{19}$m$^{-2}$ using lowest reported value for polycrystalline Cobalt films[40] (α$_0$ =5 10$^{-3}$) and the Gilbert damping of *S* obtained using broadband FMR (α =7 10$^{-3}$, see supplementary information).

In order to estimate the MoS$_2$ spin to charge conversion efficiency, the most relevant theoretical framework is that of the inverse RE effect (iREE), interestingly enough 2D TMDs have both characteristics that allow for significant Rashba SOC, 1) the lack of inversion symmetry and 2) a large Rashba splitting. According to first principle studies, SOC for MoS$_2$ can reach 150 meV for the valence band and is of few tenth of meV for the conduction band[2]. The figure of merit of iREE is given by the iREE length ($\lambda_{iREE} = \widetilde{J_C}/J_S$), $\widetilde{J_C}$ here is the charge current density expressed in unit of Am$^{-1}$ as we consider a 2D material. The iREE length depends on the Rashba field parameter($\alpha_R$) and the momentum relaxation time($\tau$), its theoretical expression is the $\lambda_{iREE} = \alpha_R \tau/\hbar$ [41].

In the following we will give an estimate of the lower bound of the spin to charge conversion efficiency ($\lambda_{iREE}$). In order to do so we need to estimate the upper bound for the value of the spin-current density $J_S$ generated by spin-pumping (eq. 1). The two most crucial parameters are the amplitude of $h_{rf}$, the rf-excitation field and $G_{\uparrow\downarrow}^{eff}$, the effective spin-mixing conductance. The maximum value of the rf-field $h_{rf}$ has been simulated for our coplanar waveguide at 20 dBm rf power to be 0.02 Oe ; and the maximum value for $G_{\uparrow\downarrow}^{eff}$ has been calculated earlier. The maximum spin current density $J_s$ is therefore 3.49 × 10$^3$ Am$^{-2}$ at 8 GHz. In Fig. 2c, we observe a spin to charge conversion voltage of 90 nV, and the resistance of the FM film is approximately 5 Ω. Taking into account the ~40% areal coverage of the monolayer MoS$_2$ islands we obtain $\lambda_{iREE}$ = 4.3 nm. This estimate is exceedingly conservative and any further precise evaluation of the calculation parameters can only induce a larger $\lambda_{iREE}$. Nevertheless, this value is significantly higher than λ$_{iREE}$ reported for topological states at the Ag/Bi interface (0.5 nm) [21] or α-Sn (2.1 nm)[24] reflecting the unique spin-electronic properties of the MoS$_2$ monolayers. To allow comparison with 3D systems we can define a pseudo-spin Hall angle $\theta_{SH}$ for MoS$_2$ by dividing $\lambda_{iREE}$ over the MoS$_2$ layer thickness (0.3 nm) which gives a value of 12.7. This value is 3 times larger than that reported by Mellnik *et al.*[4] for Bi$_2$Se$_3$ and 200 times larger than that of Pt. An effective spin-Hall-angle as large as 12.7 cannot be justified within the standard inverse SHE 3D picture nor by defects mediated spin-scattering. This large efficiency value is indicative of an intrinsic 2D mechanism for spin to charge conversion.

Taking advantage of the semiconducting character of MoS$_2$, we investigated the back-gate voltage (V$_{BG}$) dependence of the photo-voltage lineshape for *S*. Fig. 3a shows the voltage signal measured at zero back-gate, showing the spin to charge conversion in red triangular marks. The linewidths of spin to charge signal (peak 2, red triangles for the L component) at different back-gate voltages are plotted as red circles in Fig. 3b. For V$_{BG}$ = 10, 5 and 0V, there is no significant difference in the linewidths. For negative V$_{BG}$ we observe an increase of the linewidth; at -10 V bias, the linewidth and therefore the spin-relaxation-rate is 30% larger than the reference value at zero bias. This increase can be correlated to the non-Ohmic character of the Al\MoS$_2$ contact, negative back-gate corresponding, in our convention, to an increase of the interface conductivity. As a confirmation of the non-Ohmic character of the Al/MoS$_2$ contact, we performed resistance measurements on a MoS$_2$ flake in a field-effect-transistor like geometry. The channel resistance R$_{SD}$ dependence on V$_{BG}$ is reminiscent of an asymmetric back-to-back diode configuration (inset of Fig.3b). The increase in the linewidth of the *iREE* signal under negative V$_{BG}$ can hence be understood as an increase of the interface transparency to spin current and therefore to the effective spin-mixing conductance giving rise to larger total spin-relaxation rate.

In conclusion, our study of the FM/n-type MoS$_2$ system demonstrating efficient and tunable spin to charge conversion opens up multiple possibilities of interplay between charge and spin signals in monolayer TMDs. For p-type (hole-doped) materials, the effect might be much more prominent since TMDs possess large intrinsic SOC in the valence band (~300 meV Rashba splitting has been reported for WSe$_2$[42]). Furthermore Large motilities (up to 34000 cm$^2$/Vs) have recently been demonstrated[43]. Alternatively, spin injection via spin-pumping from out of plane polarized magnetic layers could allow for long spin lifetime in TMDs since the out-of plane spins states are topologically protected, making TMDs the possible next ultimate material for spintronics.

**Acknowledgments**

This work has been supported in part by the French Agence Nationale de la Recherche (ANR) NANOSWITI, MC acknowledges DGA for financial support. We would like to thank Dr. Jean-Marie George and Dr Henri Jaffrès for useful comments.

# Methods

## Growth and transfer process of $MoS_2$

0.3 g of ammonium heptamolybdate (AHM, Sigma-Aldrich, 431346) powder was dissolved into deionized (DI) water. A drop (6 µl) of the solution was released with micro-pipet onto quartz wafer (~5 x 5 mm$^2$) and dried at 80 °C oven for 3 min. The substrate was then placed next to target wafer (in CVD reactor) which was coated by sodium cholate solution as we described elsewhere [44]. Note that, AHM source should be located at more upstream region than where the target substrate is place (zone 1) d. 200 mg of S was also placed at another heating zone[45] (zone 2) prior to growth process. The temperature of zone 1 and 2 were ramped to 210 °C (42 °C/min) and 780 °C (78 °C/min), respectively. $MoS_2$ flakes were grown for 5 min with 500 sccm of $N_2$. After growth process, $MoS_2$ flake was carefully transferred onto $Si/SiO_2$ wafers by PMMA-assisted method with 1.0 mol KOH solution[46].

## Deposition of the ferromagnetic films

A continuous Al 3 nm/Co 10 nm/Al 3 nm/Cu 3 nm film was deposited using dc magnetron sputtering onto the $Si/SiO_2$ covered by the $MoS_2$ triangular shaped flakes (Fig 1.d). The base pressure of the chamber was $10^{-8}$ Torr. The Al, Co and Cu layers were sputtered under the current of 0.2 A at the growth rate of 1.33 Angstrom/sec, 0.125 A at 0.828 Angstrom/sec and 0.225 A at 5.633 Angstrom/sec, respectively. The films were measured for FMR and iREE as deposited, without any chemical or heat treatment.

## iREE measurements

A radiofrequency generator was feed into a coplanar wave guide (CPW) using an SMA connector, the sample was placed on top of the signal line of the CPW with the magnetic stack facing up. Dc contacts were then bonded directly on the film using Al wires. The rf signal was TTL modulated at 5 kHz and the photo-voltage was measured using an analog lock-in amplifier. Angular dependence of the signal was recorded to confirm the spin-to charge conversion of the photo-voltage. All measurements have been performed at room temperature.

## FMR measurements

Broadband FMR was measured using the field modulation technique in reflexing geometry using an SWR Anritsu autotester and 50Ω coplanar wave guide (CPW), the magnetic layer facing the CPW (flip-chip geometry). Field modulation had rms amplitude of 1 Oe or less. All measurements have been performed at room temperature.

**Figures**

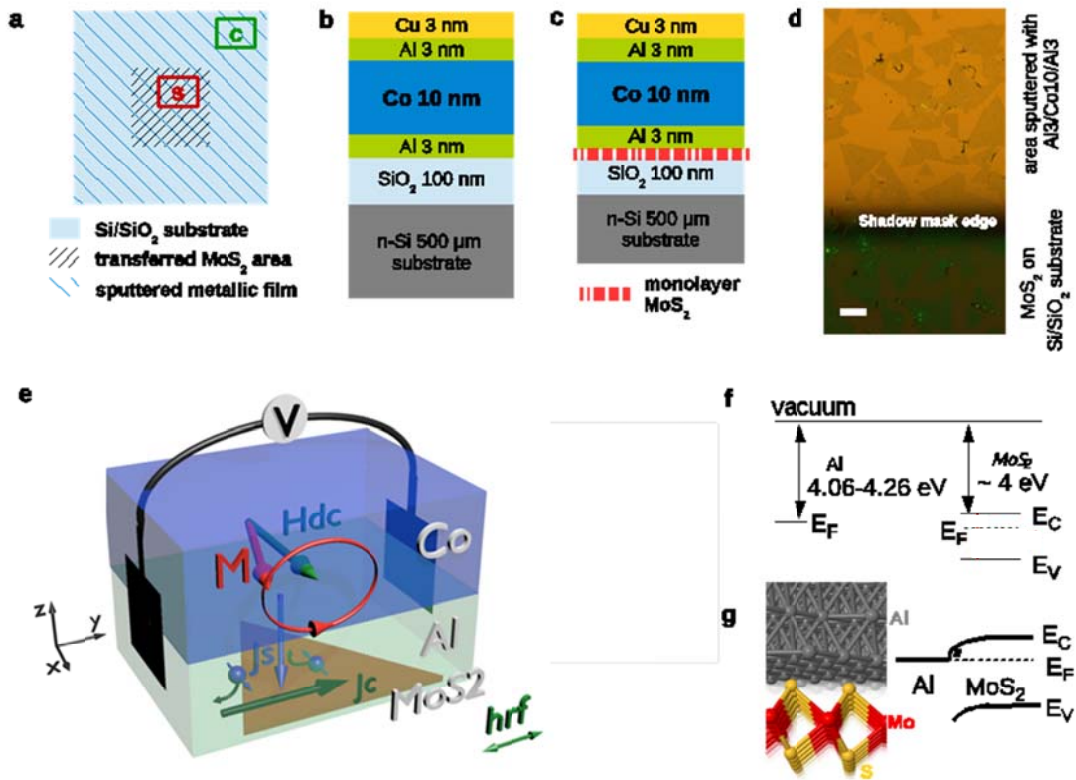

**Figure 1**: **a**, Schematic illustration of the sample (1 cm × 1 cm wafer), cut into the control sample *C* and the sample *S* with monolayer MoS$_2$. **b**, film structure of sample *C*. **c**, film structure of sample *S*, showing the monolayer MoS$_2$ between the FM stack and the Si/SiO$_2$ substrate. **d**, Optical microscope image of the monolayer MoS$_2$ after film deposition with a shadow mask, scale bar 20 μm. **e**, Sample *S* under FMR; the spin σ$_+$ is injected into the monolayer MoS$_2$ (green triangle) from the Co (FM) layer, and converted to a voltage signal V. **f**, Flat band diagram of the Al layer and the MoS$_2$ monolayer, showing the Al work function $\Phi_{Al}$ and the MoS$_2$ electron affinity $\chi_{MoS2}$. The monolayer MoS$_2$ has a bandgap of 1.29 eV and is n-type. **g**, DFT calculation of the interface atomic structure of the interface between Al and MoS$_2$.

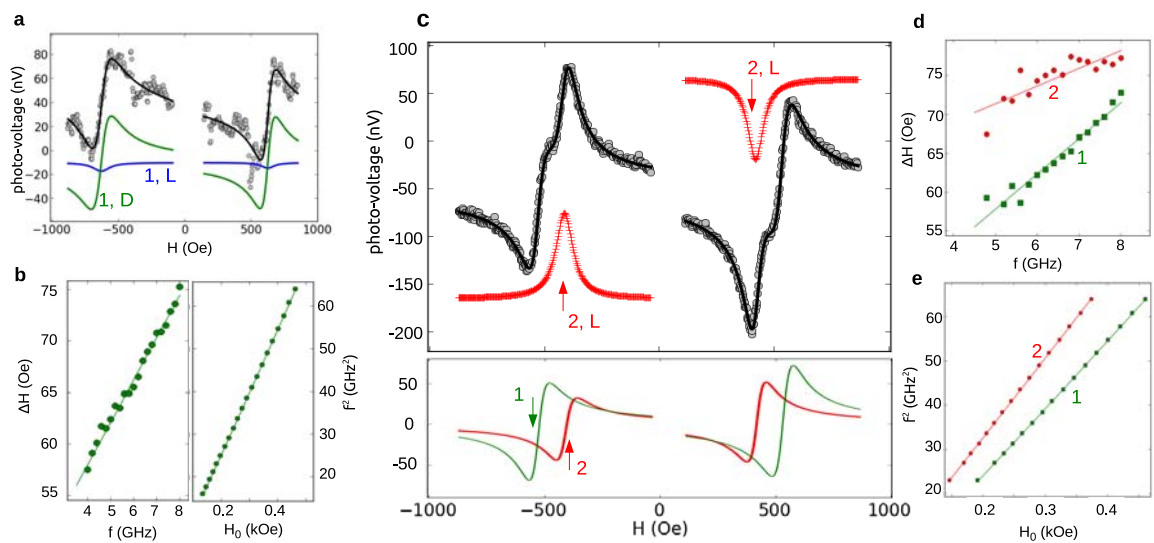

**Figure 2**: Measured photo-voltage under FMR **a**, at 8 GHz for the control sample *C* (SiO$_2$\\Al\Co\Al\Cu), the black line is the fitted lineshape; the dispersive component (1, D green line) and the Lorentzian component (1, L blue line) are shown separately. **b**, linewidth (left panel) of the peak observed in *C* as a function of frequency, showing linear dependence; relationship between the peak position and the frequency (right panel), fitted with the Kittel relation. **c**, sample *S* (SiO$_2$\\MoS$_2$\Al\Co\Al\Cu), showing two peaks; peak 1 has only AMR and PHE contributions (green line); peak 2 has an AMR and PHE contribution identical to peak 1 (red line), and an additional Lorentzian component (red cross). This signal is attributed to spin to charge conversion in MoS$_2$ flakes. **d**, linear dependence of the linewidths on frequency and **e**, Kittel relation for the two peaks in *S*. All measurements have been performed at room temperature.

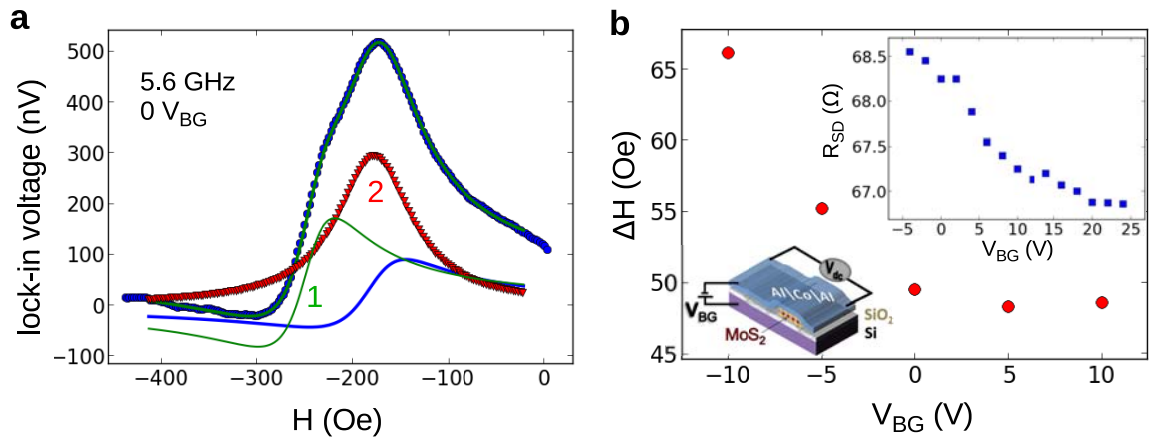

**Figure 3**: **a** Measured photo-voltage under FMR, at 5.6 GHz and zero gate voltage for sample *S* (blues circles) fitted with the cyan line. The contributions of the fitted lineshape : respectively peak 1 (green) and peak 2 (blue) that both include dispersive and Lorentzian components. The Lorentzian component of peak 2 induced by the spin to charge conversion is plotted as red triangles. **b** the back-gate dependence of peak 2 linewidth showing the increase of the spin relaxation rate when the Si\SiO$_2$\\MoS$_2$\Al\Co\Al junction is polarized toward smaller Schottky barrier height. This increase is interpreted as resulting from larger spin current being pumped to MoS$_2$. In inset is plotted the back-gate dependence of a MoS$_2$ flake resistance contacted with two Al electrodes showing the non-Ohmic character of the Al\MoS$_2$ contact.

References

1   Mak, K. F., Lee, C., Hone, J., Shan, J. & Heinz, T. F. Atomically Thin MoS_{2}: A New Direct-Gap Semiconductor. *Physical Review Letters* **105**, doi:10.1103/PhysRevLett.105.136805 (2010).
2   Zhu, Z. Y., Cheng, Y. C. & Schwingenschlögl, U. Giant spin-orbit-induced spin splitting in two-dimensional transition-metal dichalcogenide semiconductors. *Physical Review B* **84**, doi:10.1103/PhysRevB.84.153402 (2011).
3   Ochoa, H., Guinea, F. & Fal'ko, V. I. Spin memory and spin-lattice relaxation in two-dimensional hexagonal crystals. *Physical Review B* **88**, doi:10.1103/PhysRevB.88.195417 (2013).
4   Gong, K. *et al.* Electric control of spin in monolayer WSe2 field effect transistors. *Nanotechnology* **25**, doi:10.1088/0957-4484/25/43/435201 (2014).
5   Mak, K. F., He, K., Shan, J. & Heinz, T. F. Control of valley polarization in monolayer MoS2 by optical helicity. *Nature nanotechnology* **7**, 494-498, doi:10.1038/nnano.2012.96 (2012).
6   Zeng, H., Dai, J., Yao, W., Xiao, D. & Cui, X. Valley polarization in MoS2 monolayers by optical pumping. *Nature nanotechnology* **7**, 490-493, doi:10.1038/nnano.2012.95 (2012).
7   Cao, T. *et al.* Valley-selective circular dichroism of monolayer molybdenum disulphide. *Nature communications* **3**, 887, doi:10.1038/ncomms1882 (2012).
8   Mak, K. F., McGill, K. L., Park, J. & McEuen, P. L. The valley Hall effect in MoS2 transistors. *Science* **344**, 1489-1492, doi:10.1126/science.1250140 (2014).
9   Feng, W. *et al.* Intrinsic spin Hall effect in monolayers of group-VI dichalcogenides: A first-principles study. *Physical Review B* **86**, doi:10.1103/PhysRevB.86.165108 (2012).
10  Xu, X. D., Yao, W., Xiao, D. & Heinz, T. F. Spin and pseudospins in layered transition metal dichalcogenides. *Nature Physics* **10**, 343-350, doi:10.1038/nphys2942 (2014).
11  Manchon, A. Spin-orbitronics: A new moment for Berry. *Nat Phys* **10**, 340-341, doi:10.1038/nphys2957 (2014).
12  Datta, S. & Das, B. ELECTRONIC ANALOG OF THE ELECTROOPTIC MODULATOR. *Applied Physics Letters* **56**, 665-667, doi:10.1063/1.102730 (1990).
13  Rojas-Sanchez, J. C. *et al.* Spin Pumping and Inverse Spin Hall Effect in Platinum: The Essential Role of Spin-Memory Loss at Metallic Interfaces. *Physical Review Letters* **112**, doi:10.1103/PhysRevLett.112.106602 (2014).
14  Zhang, W. F., Han, W., Jiang, X., Yang, S. H. & Parkin, S. S. P. Role of transparency of platinum-ferromagnet interfaces in determining the intrinsic magnitude of the spin Hall effect. *Nature Physics* **11**, 496-+, doi:10.1038/nphys3304 (2015).
15  Pai, C. F. *et al.* Spin transfer torque devices utilizing the giant spin Hall effect of tungsten. *Applied Physics Letters* **101**, doi:10.1063/1.4753947 (2012).
16  Cubukcu, M. *et al.* Spin-orbit torque magnetization switching of a three-terminal perpendicular magnetic tunnel junction. *Applied Physics Letters* **104**, 5, doi:10.1063/1.4863407 (2014).
17  Ohshima, R. *et al.* Observation of spin-charge conversion in chemical-vapor-deposition-grown single-layer graphene. *Applied Physics Letters* **105**, 162410, doi:doi:http://dx.doi.org/10.1063/1.4893574 (2014).
18  Mendes, J. B. S. *et al.* Spin-Current to Charge-Current Conversion and Magnetoresistance in a Hybrid Structure of Graphene and Yttrium Iron Garnet. *Physical Review Letters* **115**, 226601 (2015).
19  Balakrishnan, J. *et al.* Giant spin Hall effect in graphene grown by chemical vapour deposition. *Nature communications* **5**, doi:10.1038/ncomms5748 (2014).
20  Avsar, A. *et al.* Spin–orbit proximity effect in graphene. *Nature communications* **5**, doi:10.1038/ncomms5875 (2014).

21  Sanchez, J. C. *et al.* Spin-to-charge conversion using Rashba coupling at the interface between non-magnetic materials. *Nature communications* **4**, 2944, doi:10.1038/ncomms3944 (2013).
22  Mellnik, A. R. *et al.* Spin-transfer torque generated by a topological insulator. *Nature* **511**, 449-451, doi:10.1038/nature13534 (2014).
23  Li, C. H. *et al.* Electrical detection of charge-current-induced spin polarization due to spin-momentum locking in Bi2Se3. *Nat Nano* **9**, 218-224, doi:10.1038/nnano.2014.16

http://www.nature.com/nnano/journal/v9/n3/abs/nnano.2014.16.html#supplementary-information
   (2014).
24  Rojas-Sánchez, J. C. *et al.* Spin to Charge Conversion at Room Temperature by Spin Pumping into a New Type of Topological Insulator: $\ensuremath{\alpha}$-Sn Films. *Physical Review Letters* **116**, 096602 (2016).
25  Shiomi, Y. *et al.* Spin-Electricity Conversion Induced by Spin Injection into Topological Insulators. *Physical Review Letters* **113**, 196601 (2014).
26  Sangiao, S. *et al.* Control of the spin to charge conversion using the inverse Rashba-Edelstein effect. *Applied Physics Letters* **106**, 172403, doi:doi:http://dx.doi.org/10.1063/1.4919129 (2015).
27  Zhang, W. *et al.* Research Update: Spin transfer torques in permalloy on monolayer MoS2. *APL Mater.* **4**, 032302, doi:doi:http://dx.doi.org/10.1063/1.4943076 (2016).
28  Hirsch, J. E. Spin Hall effect. *Physical Review Letters* **83**, 1834-1837, doi:10.1103/PhysRevLett.83.1834 (1999).
29  Tserkovnyak, Y., Brataas, A. & Bauer, G. E. W. Spin pumping and magnetization dynamics in metallic multilayers. *Physical Review B* **66**, doi:10.1103/PhysRevB.66.224403 (2002).
30  Tserkovnyak, Y., Brataas, A. & Bauer, G. E. W. Enhanced Gilber damping in thin ferromagnetic films. *Physical Review Letters* **88**, doi:10.1103/PhysRevLett.88.117601 (2002).
31  Chen, J. R. *et al.* Control of Schottky barriers in single layer MoS2 transistors with ferromagnetic contacts. *Nano letters* **13**, 3106-3110, doi:10.1021/nl4010157 (2013).
32  Dankert, A., Langouche, L., Kamalakar, M. V. & Dash, S. P. High-Performance Molybdenum Disulfide Field-Effect Transistors with Spin Tunnel Contacts. *ACS Nano* **8**, 476-482, doi:10.1021/nn404961e (2014).
33  Ando, K. *et al.* Electrically tunable spin injector free from the impedance mismatch problem. *Nature materials* **10**, 655-659, doi:10.1038/nmat3052 (2011).
34  Rashba, E. I. Theory of electrical spin injection: Tunnel contacts as a solution of the conductivity mismatch problem. *Physical Review B* **62**, R16267-R16270, doi:10.1103/PhysRevB.62.R16267 (2000).
35  Fert, A. & Jaffres, H. Conditions for efficient spin injection from a ferromagnetic metal into a semiconductor. *Physical Review B* **64** (2001).
36  Garello, K. *et al.* Symmetry and magnitude of spin-orbit torques in ferromagnetic heterostructures. *Nature nanotechnology* **8**, 587-593, doi:10.1038/nnano.2013.145 (2013).
37  Harder, M., Cao, Z. X., Gui, Y. S., Fan, X. L. & Hu, C. M. Analysis of the line shape of electrically detected ferromagnetic resonance. *Physical Review B* **84**, 054423 (2011).
38  Obstbaum, M. *et al.* Inverse spin Hall effect in Ni81Fe19/normal-metal bilayers. *Physical Review B* **89**, doi:10.1103/PhysRevB.89.060407 (2014).
39  Das, S., Chen, H.-Y., Penumatcha, A. V. & Appenzeller, J. High Performance Multilayer MoS2 Transistors with Scandium Contacts. *Nano letters* **13**, 100-105, doi:10.1021/nl303583v (2013).
40  Schreiber, F., Pflaum, J., Frait, Z., Muhge, T. & Pelzl, J. GILBERT DAMPING AND G-FACTOR IN FEXCO1-X ALLOY-FILMS. *Solid State Commun.* **93**, 965-968, doi:10.1016/0038-1098(94)00906-6 (1995).

| | |
|---|---|
| 41 | Shen, K., Vignale, G. & Raimondi, R. Microscopic Theory of the Inverse Edelstein Effect. *Physical Review Letters* **112**, 096601 (2014). |
| 42 | Yuan, H. *et al.* Generation and electric control of spin-valley-coupled circular photogalvanic current in WSe2. *Nature nanotechnology* **9**, 851-857, doi:10.1038/nnano.2014.183 (2014). |
| 43 | Cui, X. *et al.* Multi-terminal transport measurements of MoS2 using a van der Waals heterostructure device platform. *Nat Nano* **10**, 534-540, doi:10.1038/nnano.2015.70 http://www.nature.com/nnano/journal/v10/n6/abs/nnano.2015.70.html#supplementary-information (2015). |
| 44 | Han, G. H. *et al.* Seeded growth of highly crystalline molybdenum disulphide monolayers at controlled locations. *Nature communications* **6**, doi:10.1038/ncomms7128 (2015). |
| 45 | Zhang, Y. *et al.* Controlled Growth of High-Quality Monolayer WS2 Layers on Sapphire and Imaging Its Grain Boundary. *Acs Nano* **7**, 8963-8971, doi:10.1021/nn403454e (2013). |
| 46 | van der Zande, A. M. *et al.* Grains and grain boundaries in highly crystalline monolayer molybdenum disulphide. *Nature materials* **12**, 554-561, doi:10.1038/nmat3633 (2013). |

# Supplementary information

Flip-chip broadband FMR (see methods) has been performed to get the effective saturation magnetization $4\pi M_{eff}$ and the damping parameter α, in the frequency range of 5-20 GHz with step of 1 GHz on *C* and *S*. The results are demonstrated in Figure A. Fig. A.a and b are the FMR signal at 12 GHz for *C* and *S*, respectively. A single peak is observed in *C* but an additional peak (2, red arrow) in *S*, consistent with the voltage measurements in Fig. 2. Since the frequency and magnetic field range are much wider in this FMR setup, it is possible to distinguish the two peaks explicitly at high enough frequencies. The fittings of the linewidths and the resonance position are presented in Fig. 3c and d for *C*, and in Fig. 3e and f for *S*. The linewidth and resonance-field dependence on frequency for C and peak 1 for S (green) are almost identical which confirms our assumption that the two-peaks feature in sample *S* is due to the difference in seed layer for the metallic stack growth.

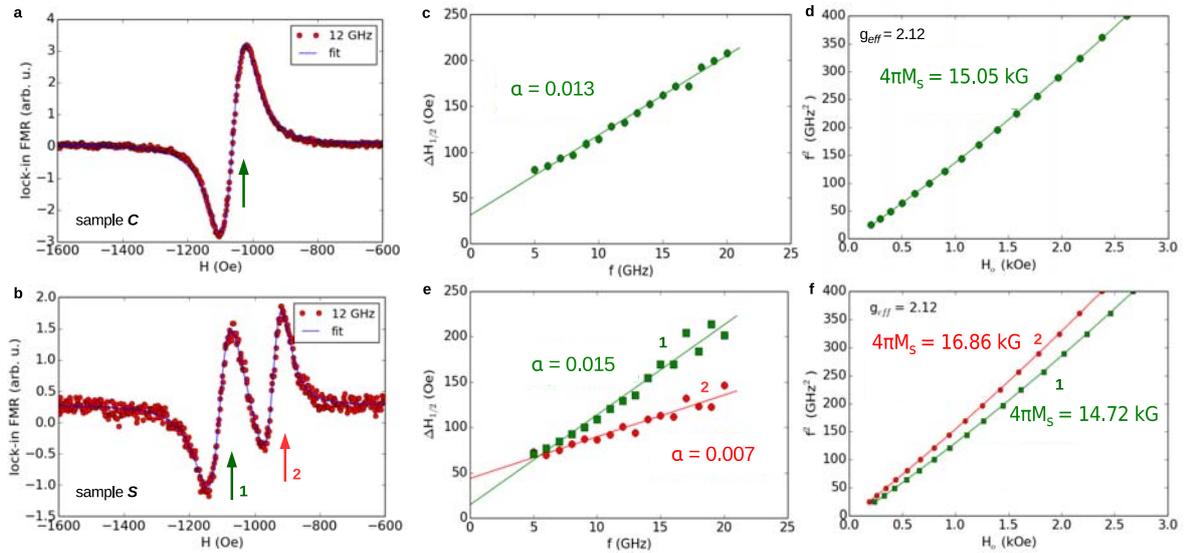

**Figure A**: FMR signal at 12 GHz for **a**, the control sample *C* and **b**, sample *S* showing two peaks. **c** (**e**), linewidth as a function of frequency to extract the damping parameter α for *C* (*S*, peak 1 (green squares) and peak 2 (red circles)). **d** (**f**), Kittel fit to extract $4\pi M_s$ for *C* (*S*, peak 1 (green squares) and peak 2 (red circles)).